# The shear viscosity and self-diffusion in ordinary water

*Viktor N. Makhlaichuk, Nikolay P. Malomuzh*

Dept. of Theoretical Physics, Odessa National University, 2 Dvoryanskaja str., Odessa, 65082, Ukraine, interaktiv@ukr.net

The paper is devoted to detailed analysis of two important transport processes – the kinematic shear viscosity and the self-diffusion – for all states of liquid water from the triple to critical points. Our approach to the shear viscosity is grounded on friction effects between two nearest molecular layers shifting relative to each other. In this relation the nature of the shear viscosity in water is fully similar to that in argon. The contribution, caused by interlayer displacements of molecules, is assumed to be negligibly small. The behavior of the kinematic shear viscosities of water is investigated in details in two characteristic directions: the vapor-liquid coexistence curve and isotherms. As for argon it is assumed that the self-diffusion in water is formed by two main contributions, caused by the molecular transport by nano-scale vortex hydrodynamic modes and collective intermixing of molecules on molecular scales. The mechanism of the second type is with good accuracy described by the Einstein formula with hard-core radius of a molecule determined from analysis of the shear viscosity of water.

## Introduction

Water demonstrates many unusual properties distincting it from argon and other argon-like liquids. It is connected with existence of H-bonds, which stimulate the clusterization processes [1-7]. However, clusters have finite life time close to that for H-bonds. Let $\tau_H$ be the life time for H-bonds [8-10]. In fact $\tau_H$ is proportional to the period of rotational motion of water molecules and it can be satisfactory estimated with the help of the dipole relaxation [1,11, 12-14].

*Dipole relaxation time in water*

The temperature dependence of the dipole relaxation time $\tau_d$ is presented in Fig.1.

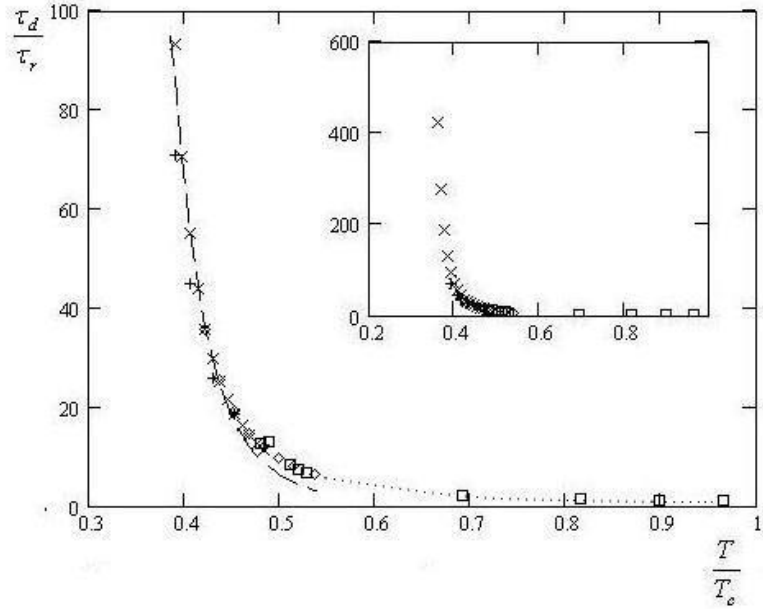

Fig.1. Temperature dependence of the dimensionless dipole relaxation time $\tilde{\tau}_d(t) = \tau_d(t)/\tau_r$, where $\tau_d$ is taken from: + - [1], quadrates – [12], crosses – [13], rhombs – [14], $\tau_r$ is the rotation period for an isolated water molecule: $\tau_r = 2\pi/(k_B T/I)^{1/2} \sim 5 \cdot 10^{-13}\,s$. Dots present interpolation values of $\tau_d$.

As we can see, the behavior of $\tilde{\tau}_d$ is essentially different in two temperature intervals: 1) $0.35 < \tilde{t} < 0.5$, $\tilde{t} = T/T_c$, where $\tilde{\tau}_d(t) >(\gg)1$ and 2) $0.5 < \tilde{t} < 0.95$, where $\tilde{\tau}_d(t) \sim 1$. In this case the exponential increase of $\tau_d$ is compatible with like-jerk rotation of water molecules and it is to lead to suitable contribution to its shear viscosity

It is necessary to note that the temperature dependence of $\tilde{\tau}_d(t)$ in the first temperature interval is satisfactory approximated by the exponential function:

$$\tilde{\tau}_d = \tilde{\tau}_d^{(0)} \exp(\varepsilon_H / \tilde{t}), \quad \tilde{\tau}_d^{(0)} = 5{,}1 \cdot 10^{-4}, \quad \varepsilon_H = 4{,}71, \qquad (1)$$

where $\varepsilon_H = E_H/k_B T_c$ and the activation energy $E_H$ for rotational motion practically coincide with the H-bonding energy [1,15-19]. This contribution is presented by the dashed line in the Fig.1. By order of magnitude: $\tau_H \sim \tau_d$.

In the temperature interval: $0.5 < \tilde{t} < 0.95$ water molecules rotate quasi-freely, therefore we can conclude that thermodynamic and many kinetic properties of water are determined by the averaged potentials having argon-like character [20].

For $\tilde{t} < 0.5$ the cluster vibration times $\tau_{cl} < \tau_d$ therefore some properties of water are caused by thermal excitations of clusters. The entropy of water and its heat capacity are the typical examples of this kind. At the same time the equation of state for water as well as its static shear viscosity and self-diffusion of molecules are formed by comparatively slow processes for which the characteristic times $\tau_{conf} > \tau_d$ ($\tau_{conf}$ is the life time for typical molecular configurations similar to those for argon). In this case, properties of water are determined by the averaged interparticle potentials [20,21].

Since the averaged interparticle potential in water is similar to that for argon [20,21] we expect that the physical nature and main properties of their static kinematic shear viscosities are similar to each other. In order to reinforce this fact let us consider their dependences on the specific volumes.

*Argon-like behaviour of the kinematic shear viscosity of water*

For this aim we apply to variables using for argon [16,22,23]: $\tilde{\nu} = \nu/\nu_R$ and the combination: $\tilde{\upsilon} = \dfrac{\upsilon - \upsilon_{tr}}{\upsilon_R - \upsilon_{tr}}$, where $\nu$ is the kinematic shear viscosity, $\upsilon$ is the specific volume per molecule. The symbols "$tr$" and "$R$" denote the values of corresponding quantities on the coexistence curve at the triple and regularization points: $\upsilon_{tr} = \upsilon(T_{tr})$, $\nu_R = \nu(T_R)$ and $\upsilon_R = \upsilon(T_R)$. The regularization temperature $T_R$ is the characteristic one dividing water states on two intervals, the properties of which are determined by thermodynamic and scale-invariant fluctuations, i.e. it is identical to the Ginsburg temperature [24].

The comparative behavior of $\tilde{\nu}$ as a function of $\tilde{\upsilon}$ for water and argon is presented in the Fig.2.

We can see that values of normalized kinematic shear viscosities for argon and water become different from each other only for $\tilde{\upsilon} < 0.3$ that corresponds to $T/T_c < 0.5$, i.e. it is observed in the region where the dipole relaxation time in water becomes exponentially increasing. .

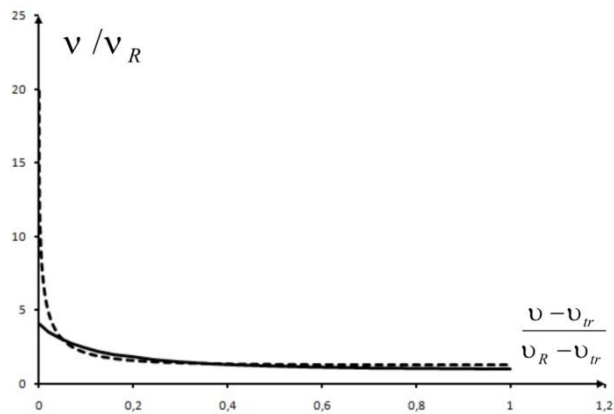

Fig.2. Values of $\tilde{v}$ as a function of $\tilde{v}$ for water and argon: solid line corresponds to argon, dashed one – to water.

The usage of simpler variables: $\tilde{v} = v/v_R$, $\tilde{t} = T/T_c$ and $\tilde{\upsilon} = \dfrac{\upsilon}{\upsilon_R}$, characteristic for the similarity principle [21,22,25,26], does not lead to so impressive similarity.

*Rotational motion of small molecular group*

Small displacements of molecules in liquids, locked in their cages, can be connected with rotations of small molecular groups on smallish angles ( see Fig.3(a)) .

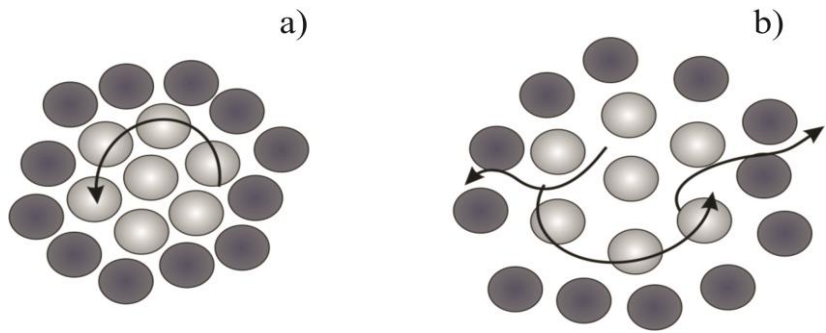

Fig.3. Schematic motion of particles near the melting or triple points (a) and near the critical one (b).

Jump-like shifts are possible too, but their contributions to the self-diffusion are expected to be negligibly small, since fluctuational voids are considerably smaller in comparison with molecular volume. Only near the critical point the through motion of molecules becomes possible (Fig.3(b)). Note that similar situation is also characteristic for movements of people in a dense crowd (here the jump-like shifts are absent!).

Thus 1) the thermal drifting of water molecules is similar to that in argon and in both these cases it has no activation character and 2) intermixing of molecules is connected with local rotational displacements of small molecular groups.

This mechanism of the self-diffusion had been firstly formulated in [27,28]. Below we show that the contribution of the local randomization to the self-diffusion coefficient is described by the Einstein formula with fixed value of a particle radius.

The present work is devoted to detailed description of the kinematic shear viscosity and self-diffusion of molecules in water. The special attention is focused on their similarity to those for argon. The important value is attached to the determination of effective radii for water molecules. This question is quite essential to guarantee the self-consistent reproduction of the kinematic shear viscosity and self-diffusion coefficient of water. We will discuss the values of water molecule radii, obtained with the help of 1) the averaged interparticle potential; 2) the excluded volume, determining the kinematic shear viscosity for water and 3) the binary correlation function. We will 1) construct the expression for the kinematic shear viscosity allowing to describe its behaviour in all directions and 2) take into account that the self-diffusion coefficient for water molecules should be considered as the sum of two contributions:

$$D_s = D_c + D_r. \qquad (3)$$

The first of them is caused by vortex hydrodynamic modes of nano-scales and the second – by the intermixing on molecular scales.

## 2. Kinematic shear viscosity of water

In this Section our main attention will be focused on the behaviour of the kinematic shear viscosity of liquid water on 1) the vapour-liquid coexistence curve and 2) isotherms, where the shear viscosity is considered as a function of pressure. In both cases we apply to the shear viscosity theory developed in [22].

*a) Kinematic shear viscosity of argon*

We begin our consideration from the kinematic shear viscosity of the simplest liquid – argon. In the following we will take into account that according to [15,16] water belongs to the class of argon-like liquids.

In accordance with [20] the kinematic shear viscosity of liquid argon on its coexistence curve is described by the formula:

$$\nu(\upsilon,t) = \nu_{tr} \frac{(\upsilon_{tr} - \upsilon_0)^{1/3}}{(\upsilon - \upsilon_0)^{1/3}}, \qquad (4)$$

where $v_{tr}$, $\upsilon_{tr}$ are the kinematic shear viscosity and specific volume per molecule at the triple point and $\upsilon_0$ is the excluded volume. With high accuracy

$$\upsilon_0 \approx \upsilon_{tr} \quad \text{and} \quad \upsilon_{tr} > \upsilon_0. \tag{5}$$

The numerical values of all parameters, introducing to (4), are equal to:

$$v_{tr} = 0.2081 \cdot 10^{-2}\, cm^2/s, \quad \upsilon_{tr} = 47.05\, \text{Å}^3, \quad \upsilon_0 = 46.10\, \text{Å}^3 \tag{6}$$

It is necessary to note, that the formula (4) differs from that in [20] by another character of normalization, here we normalize the specific volume and kinematic shear viscosity on their values at the triple point. Both these possibilities are fully equivalent. It is very important that the formula (4) describes successfully the kinematic shear viscosities of all atomic liquids (noble gases) and those low-molecular liquids for which the averaged inter-particle potentials have argon-like form, i.e. they are described by formulas of the Lennard-Jones type [20].

The formula (4) is also correct for the description of the pressure dependence for the kinematic shear viscosity of argon on isotherms (see Fig.4). As we can see the differences between calculated and experimental data [29] do not exceed the experimental error.

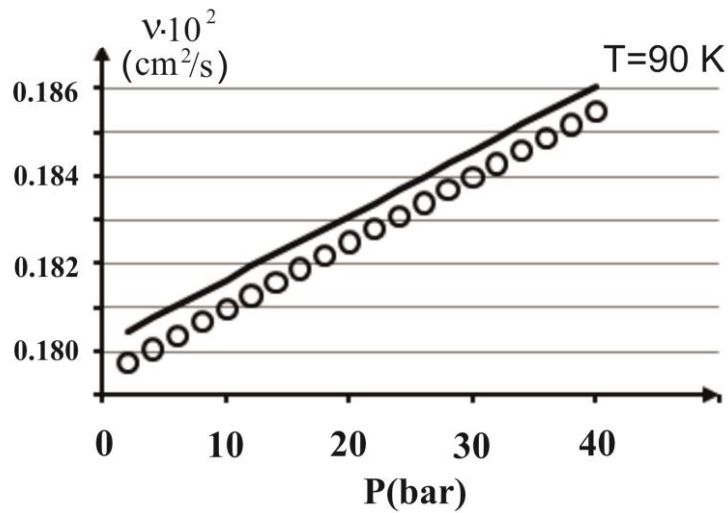

Fig.4. The kinematic shear viscosity of argon vs. pressure at $T = 90\, K$.

It is necessary to stress that the excluded volume of argon: $\upsilon_0^{(Ar)}(p) = 43.62\, Å^3$ on the isotherm $T = 90\, K$ is partly less than $\upsilon_0^{(Ar)} = 46.1\, Å^3$ on the coexistence curve.

Water belongs also to this type of liquids although this seems to be very surprising.

*b) Kinematic shear viscosity of water on its vapour-liquid coexistence curve*

The argon-like behavior of the inter-particle potentials and kinematic shear viscosity becomes clear if we take into account the rotational motion of water molecules. In this case, water properties, caused by translational molecular motions, change slower in comparison with contributions of rotational motions. Therefore the coexistence curves of water and argon are similar with quite satisfactory accuracy [20,21], so the static values of the kinematic shear viscosity for water are to be argon-like.

As a consequence, we expect that the kinematic shear viscosity of water is described by the formula:

$$v_w(\upsilon,t) = v_{tr} \frac{(\upsilon_{tr} - \upsilon_0(t))^{1/3}}{(\upsilon - \upsilon_0(t))^{1/3}}, \qquad (7)$$

where

$$v_{tr} = 1.7916 \cdot 10^{-2} \, cm^2/s, \quad \upsilon_{tr} = 29{,}8863 \, \text{Å}^3. \qquad (8)$$

It differs from that for argon by weak temperature dependence of the excluded volume: $\upsilon_0(t)$. At that, water molecules can be considered as spherical particles in consequence of their rotational motion. It is considerably expressed at $t > t_H$, $t_H \approx 1.2$ (see [30]), therefore for the temperature interval $t_H < t < t_c$, $t_c = T_c/T_{tr} \approx 2.4$, we expect that $\upsilon_0(t)$ should be constant. Determining it by the least square method we obtain:

$$\upsilon_0 = 29.8831 \, \text{Å}^3 \qquad (9)$$

The agreement of experimental data with those calculated according to (7) is demonstrated in Fig.5.

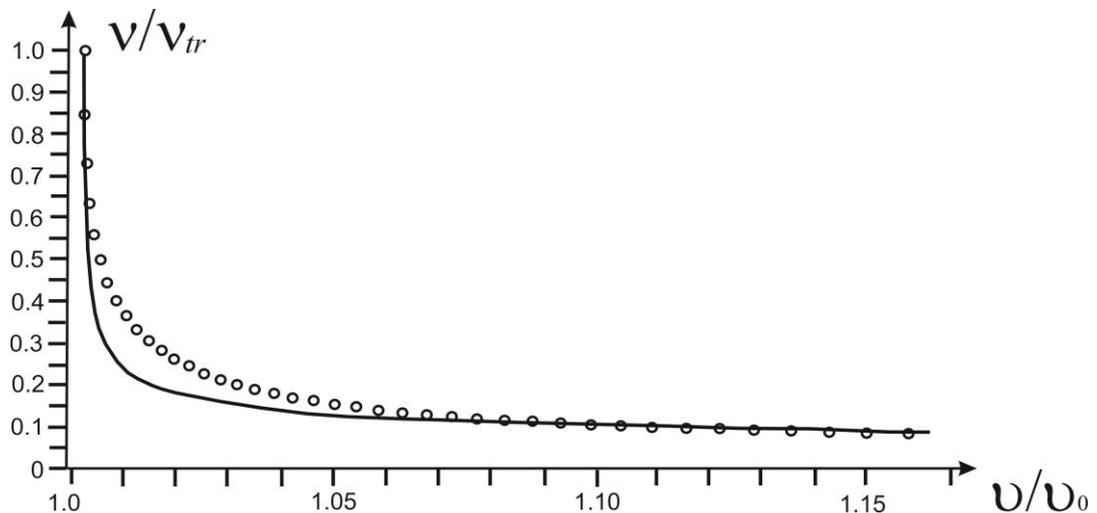

Fig.5. Dependence $v/v_{tr}$ vs. $\upsilon/\upsilon_0$, where $\upsilon_0$ is given by (9), for ordinary water: points – experimental values, solid line – calculated according to (7).

Within the temperature interval: $1 < t < t_H$ the rotation of molecules becomes essentially slower [12-14] and the influence of H-bonds leads to the weak dependence on temperature. The immediate account of this influence is rather difficult, therefore here we restrict ourselves by experimental estimate (see Fig.5). In accordance with (7) $\upsilon_0(t)$ satisfies the equation:

$$\upsilon_0(t) = \frac{\upsilon_{tr} - (v(t)/v_{tr})^3 \upsilon(t)}{1 - (v(t)/v_{tr})^3}, \qquad (10)$$

leading to the curve in Fig.6.

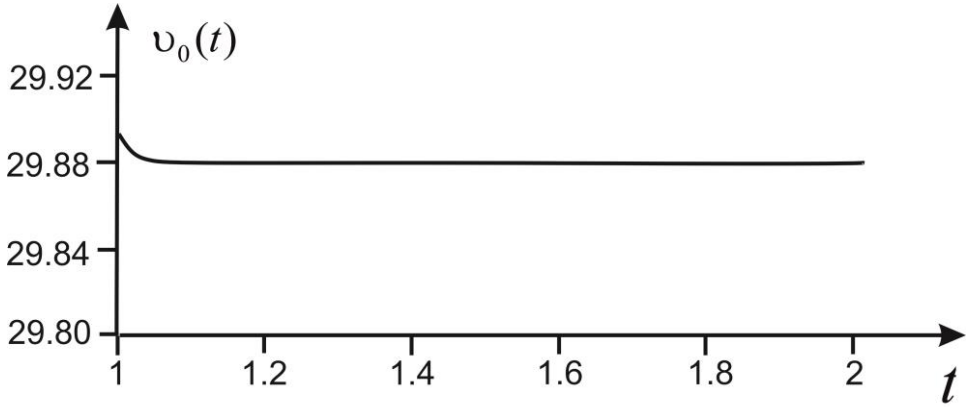

Fig.6. The excluded volume $\upsilon_0(t)$ vs. temperature

As we can see the little deviations from $\upsilon_0 = 29.8831$ $A^3$ are observed only near the triple point and they do not exceed 3% (the accuracy of experimental data [31] is about 1%). At that, at its narrow vicinity the profile of $\upsilon_0(t)$ resembles the behavior of water density on the left of its minimum at $4^0 C$.

*c) Radius of a water molecule following from the excluded volume*

It is natural to assume that the effective radius of a water molecule [23] is determined by the expression:

$$r^{(w)}(t) = \left(\frac{3}{16\pi}\upsilon_0(t)\right)^{1/3}, \qquad (11)$$

according to the theory of virial expansions. The corresponding temperature dependence of $r^{(w)}(t)$ is presented in Fig.7.

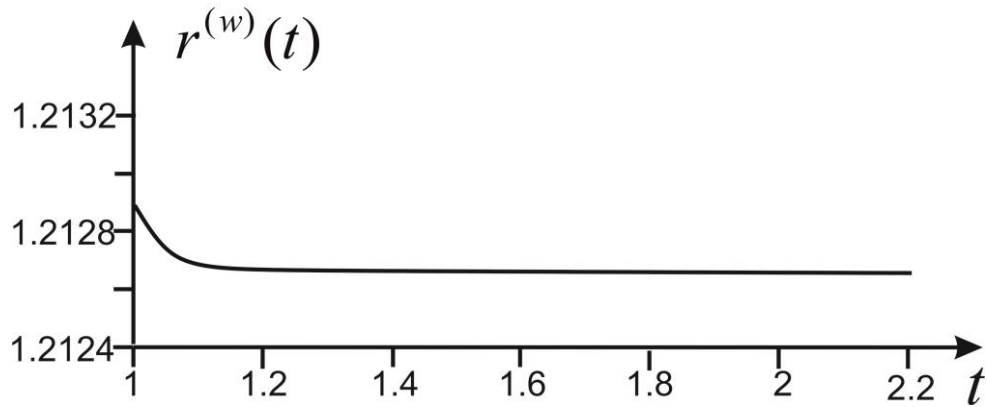

Fig.7. Temperature dependence of $r^{(w)}(t)$ for ordinary water

It is necessary to note that the values of the excluded volumes, determined from the kinematic shear viscosity and van der Waals equation of state, differ from each other [32]. The first of them is determined mainly by the repulsive part of interparticle potential while the second – by the sum of repulsive and attractive ones. This implies that $\upsilon_0(v) < \upsilon_0(vdW)$. It is necessary to stress that the accuracy of determination of $\upsilon_0(t)$ and $r^{(w)}(t)$ according to the kinematic shear viscosity is essentially higher.

*d) Kinematic shear viscosity of water on isotherms*

In principle the general character of dependence of the kinematic shear viscosity on the normalized temperature $t = T/T_{tr}$ and specific volume per particle $\upsilon$ remains the same. However we should take into account the dependence of the key parameters entering to (7) on pressure:

$$v_w(t,p) = v_m(p) \frac{(\upsilon_m(p) - \upsilon_0(p))^{1/3}}{(\upsilon(t,p) - \upsilon_0(p))^{1/3}}, \qquad (12)$$

where $\upsilon_m(p)$ is the specific volume of water at the melting temperature $t_m(p)$ corresponding to pressure $p$, $t = T/T_m(p)$ is the redefined temperature. At that, $t_m(p_{tr}) = 1$, $\upsilon_m(p_{tr}) = \upsilon_{tr}$.

The dependence of the specific volume on temperature and pressure are different within the range:

$$1 < t < 1.2, \quad p_{tr} < p < 150\,MPa, \tag{13}$$

and outside it. Within this range, the specific volume as a function of temperature decreases on isobars that also leads to the decrease of the kinematic shear viscosity. Outside it, the standard behavior occurs, i.e. $\upsilon$ decreases with the increase of pressure and $\nu$ increases with pressure.

a) Let us consider the pressure dependence of the kinematic shear viscosity for $p > 150\,MPa$, where we can apply to the Tait equation [33]:

$$\frac{\zeta_0 - \upsilon}{\zeta_0} = \frac{A(p - p_0)}{B + p - p_0},$$

where $A$, $B$ and $\zeta_0, p_0$ are constants. In accordance with [34]

$$A = 0.14, \quad B = 276.3\,MPa,$$

The values $\zeta_0 = 29.8863\,\text{Å}^3$, $p_0 = 0.1\,MPa$ are suggested to be equal to the specific volume at the triple point and atmospheric pressure.

The excluded volume $\upsilon_0$ can be estimated as the limit value of $\upsilon$ for $p - p_0 >> (>)B$. From the Tait equation we get:

$$\frac{\upsilon_0}{\zeta_0} = 1 - A.$$

As a result,

$$\frac{\upsilon(p,t) - \upsilon_0(p)}{\zeta_0} = \frac{AB}{B + p - p_0},$$

and

$$v_w(t, p) = v_m(p) \frac{(\upsilon_m(p) - \upsilon_0(p))^{1/3}}{(AB\zeta_0)^{1/3}} (B + p - p_0)^{1/3}. \tag{14}$$

In order to find the difference $\upsilon_m(p) - \upsilon_0(p)$ in accordance with the said above at larger pressure we can take the value $\upsilon_m(p)$ instead of $\upsilon_0(p)$. Here we will use the estimate: $\upsilon_m(p) - \upsilon_0(p) = \upsilon_m(p) - \upsilon_m(2p)$, taking $200\,MPa < p < 400\,MPa$. It is surprising that this difference takes the same value at $0.1\,MPa, < p < 150\,MPa$. The value of $v_m(p) = 1.523 \cdot 10^{-2}\,cm/s^2$ is supposed to be equal to the shear viscosity at $T = 278\,K$ and $p = 1\,MPa$.

The comparison of shear viscosity values, calculated according to (14), and experimental data is presented in Fig.8.

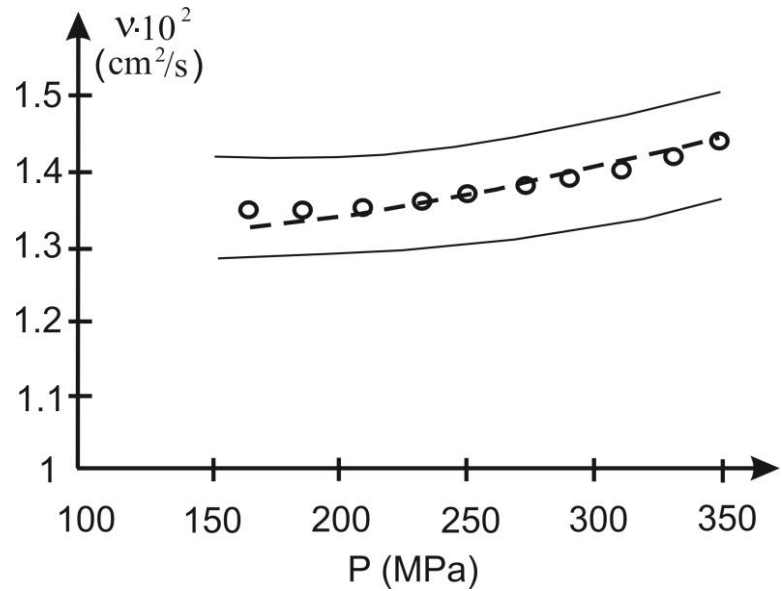

Fig.8. The kinematic shear viscosity of water as a function of pressure on the isotherm $T = 283\,K$: open circles – experimental data from [29], dashed line – calculated according to (14), thin black lines note the corridor of experimental errors ($5\,\%$).

As we can see, the applicability regions for the Tait equation and standard dependence of the kinematic shear viscosity on pressure are consistent with each other for $p > 150\,MPa$.

b) Now, let us consider the peculiarities of the kinematic shear viscosity within the range given by inequalities (13) . In this case the values of $\upsilon_m(p)$ are determined by the experimental data [29]. The values of $\upsilon_0(p)$ are determined by the temperature dependences of the kinematic shear viscosity on each isobar. As a result we get on the isotherm $T = 283\,K$:

$$\upsilon_m(p) - \upsilon_0(p) = 0.0161 + 0.00021 \cdot p,$$

where $p$ is measured in MPa. Here we take into account that the difference $\upsilon_m(p) - \upsilon_0(p)$ changes in small limits and therefore it can be approximated by linear function of pressure. Using experimental values of $\upsilon(t, p)$ on each isotherm and $\upsilon_0(p)$, determined above, on the same isotherm we get:

$$\upsilon(t, p) - \upsilon_0(p) = 0.01581 + 0.00019 \cdot p.$$

The comparison of theoretical values of the kinematic shear viscosity with corresponding experimental data is presented in Fig.9.

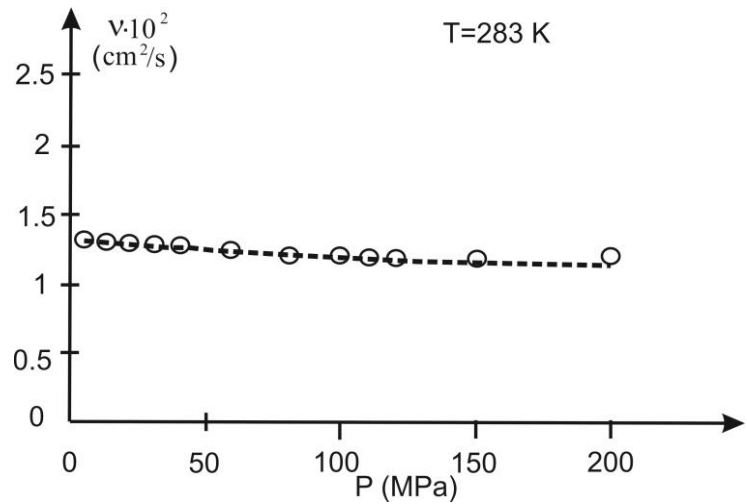

Fig.9. The kinematic shear viscosity of water vs. pressure on the isotherm $T = 283\,K$: open circles – experimental data, dashed line – calculated according to (12) and the values of nominator and denominator given above.

As a matter of fact, for $T_m(p) < T_{tr}$ and $p > p_{tr}$ the situation does not change.

## Other estimates of molecular sizes

There are several approaches for the determination of molecular sizes based on using of: 1) the van der Waals EoS; 2) the kinematic shear viscosity; 3) the structural factor and 4) the averaged interparticle potentials. Let us consider the radii of a water molecule obtained with their help.

Values of the excluded volumes, which are used for the determination of the molecular radii according to (11), are presented in the Table 1. In it $\upsilon_0(\nu)$ corresponds to the kinematic shear viscosity and $\upsilon_0(vdW)$ - to the vdW EoS. As we can see, in all cases the inequality:

$$\upsilon_0(vdW) > \upsilon_0(\nu) \approx \upsilon_{tr} \qquad (12)$$

is carried out. Here the values of $\upsilon_0(vdW)$ are determined by vapour-like phases. Another inequality

$$\upsilon_{tr} > \upsilon_0(\nu), \qquad (13)$$

also takes place, excepting benzene (B) and nitrobenzene (NB).

Table 1. Values of the parameters: $\upsilon_0(v)$, $\upsilon_0(vdW)$ and $\upsilon_{tr}$ [32]

|         | $\upsilon_{tr}$ | $\upsilon_0(v)$ | $\upsilon_0(vdW)$ |
|---------|--------|--------|--------|
| $Ar$    | 47,05  | 46,50  | 53,47  |
| $Ne$    | 26,93  | 26,04  | 28,19  |
| $Kr$    | 56,91  | 56,16  | 66,13  |
| $H_2O$  | 29,89  | 29,88  | 50,70  |
| $N_2$   | 53,86  | 52,62  | 65,00  |
| $O_2$   | 40,67  | 39,81  | 52,61  |
| $B$     | 145,09 | 153,35 | 191,69 |
| $NB$    | 167,78 | 225,34 | 255,81 |

It is necessary to note that molecules of noble gases $Ar, Ne, Kr$ have spherical shape, small molecules of water and $N_2$, $O_2$ have no spherical shape but they become quasi-spherical in consequence of their thermal rotation. As a result, their values of $\upsilon_0(v)$ and $\upsilon_{tr}$ prove to be close to each other. The peculiarities of rotation for larger-dimension non-spherical molecules B and NB lead to the violation of this condition.

Using (11) and the excluded volumes from Table 1 we find the following values of molecular radii

Table 2. Values of molecular radii determined by the kinematic shear viscosity and the van der Waals EoS.

|             | $Ar$ | $Ne$ | $Kr$ | $H_2O$ | $N_2$ | $O_2$ | $B$  | $NB$ |
|-------------|------|------|------|--------|-------|-------|------|------|
| $r_i^{(v)}$ | 1,41 | 1,16 | 1,50 | 1,21   | 1,46  | 1,33  | 2,09 | 2,38 |

| $r_i^{(vdW)}$ | 1,47 | 1,19 | 1,58 | 1,45 | 1,57 | 1,46 | 2,25 | 2,48 |

The comparison $r_{H_2O}^{(v)}$ with the hard-core radius $r_{H_2O}^{(hc)}$ of a water molecule testifies about their full coincidence:

$$r_{H_2O}^{(v)} \approx r_{H_2O}^{(hc)} \approx 1.21 \text{ A}. \tag{14}$$

Here the radius $r_{H_2O}^{(hc)} \approx \frac{l}{2}$ where $l$ the length of dashed line segment in Fig.10.

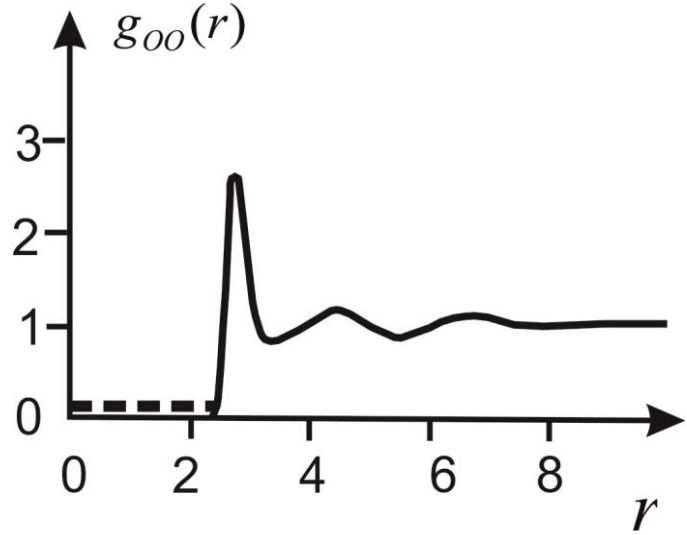

Fig.10. Binary correlation function $g_{OO}(r)$ for water at $T = 296\,K$ [35,36].

Note, that the value $r^{(w)} = 1,21\,A$ corresponding to Fig.6 is very close to (14).

It is appropriated to compare the obtained values of water molecule radius with that following from the equation

$$r_\rho^{(w)}(t) = \left( \frac{3}{16\pi} \frac{m_w}{\rho_w(t)} \right)^{1/3}, \tag{15}$$

where $m_w$ is the mass of the water molecule and $\rho_w(t)$ is the mass density of water. The numerical values of $r_\rho^{(w)}$, calculated according to (15), are presented in Fig.11. As we can see, values $r_\rho^{(w)}$ deviate from $r_{H_2O}^{(v)}$ only far away from the triple point.

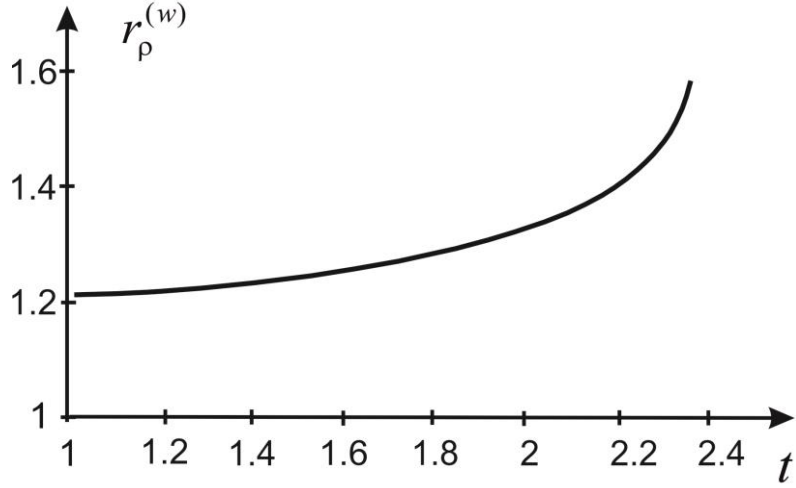

Fig.11. The water radius, calculated according to (15), vs. temperature.

*The averaged interparticle potential in liquid water and its vapor*

By the definition the averaged interparticle potential $U(r_{12})$ between two water molecules equals to (see [18,20]):

$$\exp(-\beta U(r_{12})) = \oint_{\Omega_1=4\pi} \frac{d\Omega_1}{4\pi} \oint_{\Omega_2=4\pi} \frac{d\Omega_2}{4\pi} \exp(-\beta \Phi(1,2)), \qquad (16)$$

where $\Phi(1,2) = \Phi(r_{12}, \Omega_1, \Omega_2)$ is the microscopic potential, depending on the interparticle spacing $r_{12}$ and two angles $\Omega_1, \Omega_2$ describing the spatial orientations of molecules. The potential $\Phi(1,2)$ has the structure:

$$\Phi(r,\Omega) = \Phi_R(r,\Omega) + \Phi_D(r,\Omega) + \Phi_E(r,\Omega) + \Phi_H(r,\Omega). \qquad (17)$$

where $\Phi_R(r,\Omega), \Phi_D(r,\Omega), \Phi_E(r,\Omega), \Phi_H(r,\Omega)$ are terms caused by repulsive, dispersive, electrostatic and irreducible H-bond contributions correspondingly.

The application of (16) guarantees the carrying out of the fundamental relation between the free energy and the configuration integral in two particle approximation:

$$F_2 = -T \ln Q_2, \qquad (18)$$

where

$$Q_2 = \int_V d\vec{r}_1 \int_V d\vec{r}_2 \oint_{\Omega_1=4\pi} d\Omega_1 \oint_{\Omega_2=4\pi} d\Omega_2 \exp(-\beta \Phi(1,2)) = \frac{(4\pi)^2}{2!} \int_V d\vec{r}_1 \int_V d\vec{r}_2 \exp(-\beta U(r_{12})).$$

The components; $\Phi_R(r,\Omega), \Phi_D(r,\Omega), \Phi_E(r,\Omega)$ of microscopic potential will be modeled by the corresponding terms introducing to the standard model potentials [37 - 41]:

$$\Phi_E = \begin{cases} \Phi_E^{(i)}, & i = SPC, TIPS,... \\ \Phi_M, & \text{multipole series} \end{cases} \quad (19)$$

In liquid water as well as in dense enough vapor phase, the electrostatic interaction energy between water molecules reduces in consequence of polarization effects accoding to: $\Phi_E(r,\Omega) \to \frac{1}{\varepsilon_\infty}\Phi_E(r,\Omega)$, where $\varepsilon_\infty$ is the dielectric permittivity on frequences $\omega \sim 1/\tau_r \approx 10^{-12} s$, corresponding to the rotational motion. In accordance with experimental works [42,43]: $\varepsilon_\infty = 2.4$.

Numerical values $U(r_{12})$, calculated with the help of (16), are quite satisfactory fitted by the Lennard-Jones expression:

$$U(r) = 4\varepsilon_{LJ}\left[\left(\frac{\sigma_{LJ}}{r}\right)^{12} - \left(\frac{\sigma_{LJ}}{r}\right)^{6}\right]. \quad (20)$$

The values of parameters $\varepsilon_{LJ}$ and $\sigma_{LJ}$, obtained this way for $\varepsilon_\infty = 2.4, 3$, are presented in the Tables 3 and 4 correspondingly:

Table 3. Parameters $\varepsilon_{LJ}$ and $\sigma_{LJ}$ corresponding to $\varepsilon_\infty = 2.4$

|  | SPC | TIPS | SPC/E | TIP3P |
|---|---|---|---|---|
| $\varepsilon_{LJ}/(k_B T_{tr})$ | 2.13 | 1.71 | 2.39 | 2.14 |
| $\sigma_{LJ}$ | 2.70 | 2.73 | 2.68 | 2.69 |

Table 4. Parameters $\varepsilon_{LJ}$ and $\sigma_{LJ}$ corresponding to $\varepsilon_\infty = 3$

|  | SPC | TIPS | SPC/E | TIP3P |
|---|---|---|---|---|
| $\varepsilon_{LJ}/(k_B T_{tr})$ | 1.43 | 1.42 | 1.42 | 1.42 |
| $\sigma_{LJ}$ | 2.78 | 2.78 | 2.77 | 2.77 |

As we can see, the values of $\sigma_{LJ}/2$ for potentials SPC, SPC/E, TIPS и TIP3P satisfy the inequality:

$$r_w^{(hc)} < \sigma_{LJ}/2 < l_H/2,$$

where $l_H$ is the length of an H-bond. By order of magnitude $l_H \approx l + \Delta_H$, where $\Delta_H$ is the broadening of peak of the binary correlation function (see Fig.10). For vaporous states the value of $\sigma_{LJ}$ shifted to the right and for liquid states – to the left.

### 3. The self-diffusion in water

The main attention in this Section is paid to the analysis of the self-diffusion coefficient of water molecules on i) its vapor-liquid coexistence curve and ii) isotherms, where $D_s$ is considered as a function of pressure.

*A) Self-diffusion of water molecules on the coexistence curve*

In correspondence with (3) we suppose that the self-diffusion coefficient of a water molecule is the sum of two collective components:

$$D_c = \frac{k_B T}{10\pi\eta\sqrt{\nu\tau_M}}, \qquad (21)$$

where $\tau_M$ is the Maxwell relaxation time (MRT) for high frequency viscous tensions, and the collective component of another type:

$$D_r = \frac{k_B T}{6\pi\eta r^{(v)}}, \qquad (22)$$

determined by the Einstein formula with the effective radius responsible for the shear viscosity of water.

Here it is necessary to make several important remarks. These both contributions to the self-diffusion coefficient of liquid have collective character, i.e. they are connected with simultaneous displacements of some molecular groups.

The formula (21) reflects the drift of a molecule together with a liquid particle surrounding it in the velocity field of hydrodynamic fluctuations. The size of this liquid particle is to be estimated as the correlation radius $r_c = 2\sqrt{\nu t}$ for thermal hydrodynamic fluctuations corresponding to transversal modes of the velocity field. The maximal contribution to the self-diffusion is caused by the smallest liquid particles having the size $r_L = 2\sqrt{\nu \tau_M}$. Such particles are accepted to be called the Lagrange ones. Their contribution to the self-diffusion coefficient is in details discussed in [27]. At the same time oscillating longitudinal hydrodynamic modes cannot lead to systematic transport of particles.

*a) Estimates of the MRT*

The MRT can be approximately estimated with the help of modification of the Maxwell formula:

$$\tau_M = \eta / G.$$

Since $G/\rho = c_t^2$, where $G$ is the high frequency shear modulus and $c_t$ is the transversal sound velocity in liquid water, we can write:

$$\tau_M = \nu / c_t^2 > \nu / c_l^2, \qquad (23)$$

where $c_l$ is the longitudinal sound velocity and we use the inequality: $c_l > c_t$ [23]. In fact, the inequality (23) gives us the lower limit for values of the MRT.

In the following we will assume that $c_t^2 \approx (2/3) c_l^2$ and

$$\tau_M^{(a)} \approx (3/2) \nu / c_l^2. \qquad (24)$$

Here it is assumed that the bulk modulus exceeds the shear one approximately $3/2$ times more (see [22]).

The following additional inequality is also to take place:

$$\zeta \gg (>)1, \quad \zeta = 2\sqrt{\nu \tau_M^{(a)}} / 3 r_w, \qquad (25)$$

i.e. the radius of a liquid (Lagrange) particle is to exceed the one for a water complex formed by some molecule and its nearest surrounding.

Another estimate for the MRT ($\tau_M^{(MS)}$) for water was obtained with the help of MD-method in [26, 44].

The values of $\tau_M$ calculated with the help of (24) and obtained in [44] are collected in the Table 5. The applicability region of $\tau_M^{(a)}$ and $\tau_M^{(MS)}$, determined by (23) and (25), is restricted by $t < 1.3$.

Table 5. The values of the MRT vs. temperature, satisfying to inequalities (23)

and (25)

| $t = T/T_{tr}$ | $\tau_M^{(a)} \cdot 10^{13}, s$ | $\tau_M^{(MS)} \cdot 10^{13}, s$ | $D_c / D_{exp}$ | $\zeta$ |
|---|---|---|---|---|
| 1.00 | 11.58 | 9.8 | 0.04 | 7.87 |
| 1.09 | 5.97 | 8.66 | 0.09 | 4.01 |
| 1.21 | 3.37 |  | 0.24 | 2.15 |
| 1.30 | 2.44 | 6.12 | *0.38* | 1.65 |
| 1.40 | 1.79 | 4.89 | *0.51* | 1.21 |

As we can see, the applicability region of the nanoscopic collective drift in liquid water is restricted by temperatures: $t < 1.3$. In accordance with arguments, presented in the introduction, the through translational displacement of a molecule near the triple point is impossible. Due to absence of voids of suitable volume the activation motion is also impossible. It is even impossible in solid bodies as it had been shown in [28]. Thus the main contributions to the self-diffusion arise due to systematic intermixing of particles on the molecular scale.

*b) Reasons in the favor of (22)*

Due to the dimensionless reasons this contribution has a structure:

$$D_r \sim \frac{k_B T}{\eta r_p^{(v)}},$$

where properties of liquid are characterized by the combination $\frac{k_B T}{\eta}$ and those of molecules - by their size $r_p^{(v)}$ which first of all manifest in the shear viscosity. This conclusion is the direct consequence of the molecular motion described in the subsection "Rotational motion …" and having the same character for the self-

diffusion and the formation of shear viscosity. These components satisfy the similarity relation:

$$D_r(1)/D_r(2) \sim r_p^{(v)}(2)/r_p^{(v)}(1).$$

The situation remains the same for binary mixtures too. But in this case for large enough doped particles we can use the Einstein formula for Brownian particles:

$$D_r = \frac{k_B T}{6\pi\eta r_B}.$$

Combining all these arguments we obtain the final expression (22).

*c) Comparison with experimental data*

The comparison of the theoretical and experimental results within the applicability region of the MRT, $t_{tr} < t < t_\zeta$, $t_\zeta \approx 1.2$, is presented in Fig.12. The wave line corresponds to the upper boundary for the applicability of $\tau_M^{(a)}$.

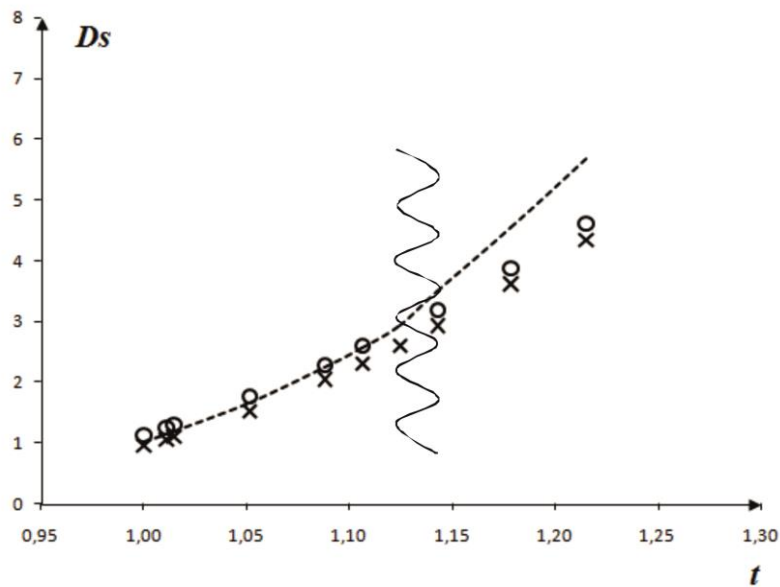

Fig.12. Temperature dependences of experimental values for the self-diffusion coefficient (open circles), component $D_r$ describing intermixing on molecular scales (crosses) and the sum, $D_r + D_c$ (dashed line), corresponding to the applicability region, $1 < t < 1.2$ of collective contribution (21). We use $\tau_M^{(a)}(t)$ and $r^{(w)} = 1.21\ A$.

As we can see from the Table 5, the ratio $D_c/D_{exp}$ changes in limits: $0.04 < D_c(t)/D_{exp}(t) < 0.24$. At that, estimates of $D_c/D_{exp}$, obtained with the help of $\tau_M^{(a)}$ and $\tau_M^{(MS)}$ differ from each other not more than $1.4 \div 1.5$ times. At $t > 1.25$ the collective transport of water molecules, described by the contribution (21), becomes impossible since the notion of a Lagrange particle loses its meaning and

$$D_s \to D_r. \qquad (23)$$

The agreement between values of the self-diffusion coefficient calculated according to (22) and their experimental data for $t > 1.3$ is presented in Fig.13.

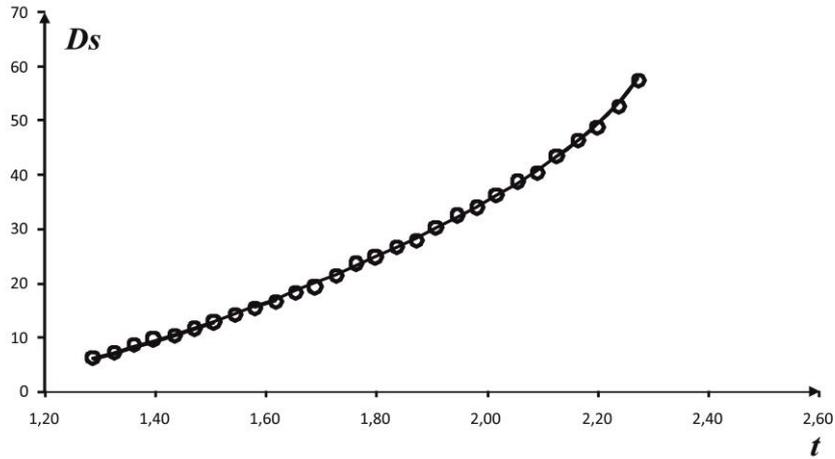

Fig.13. The self-diffusion coefficient for water molecules vs. temperature at $t > 1.25$: open circles – experimental data, filled circles – values of $D_r$. Here the value $r_{H_2O}^{(v)} = 1.21 A$ is used.

*B) The self-diffusion coefficient of water molecules on isotherms*

The general formulas, determining the behavior of the self-diffusion coefficient of water molecules remain to be the same. Here we take into account the non-monotonic dependence of the shear viscosity on pressure for all isotherms within the temperature intervals: $273 K < T_i < 323 K$ [29]. In this temperature interval the dynamic shear viscosity as a function of pressure decreases until $p < p_s$, $p_s \sim 2000\, atm$, reaches its minimum and only after this it begins to increase similarly to that of simple liquids.

Since the main contribution to the self-diffusion coefficient is determined by the contribution $D_r$ described by (19), we will suppose that namely it determines the dependence of $D_r = D_r(p, T_i)$ on pressure (see Fig.14).

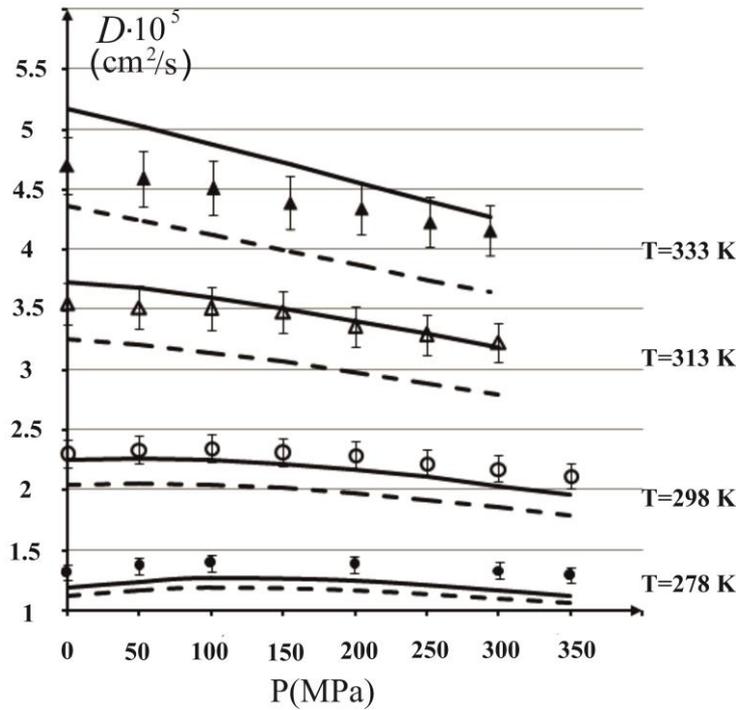

Fig.14. The self-diffusion of water molecules as a function of pressure of different temperatures. Open and filled circles and triangles - experimental values, dashed lines - values of $D_r$, solid lines - values of $D_r + D_c$. Vertical bars represent 5% experimental errors.

As we can see, the deviations between experimental values and those calculated according to (3), (21) and (22) of the self-diffusion coefficients do not exceed experimental errors (5%).

The radius of a Lagrange particle decreases with the growth of pressure: at $T = 333\,K$ it equals 7.5 A.

### Discussion of the results obtained

Our analysis of thermal motion in water leads to a conclusion that the kinematic shear viscosity and self-diffusion have argon-like character. The last is characteristic for the majority of low-molecular liquids. It means that the shear viscosity is formed by the friction effects arising because of shifting of molecular layers relative to each other. The similarity between argon and water arises due to rotational motion of water molecules. As a result, not spherical water molecules manifest themselves as spherical particles. In generalized coordinates the shear viscosity of water differs from that for argon on their coexistence curves only by weak temperature dependence of its excluded volume. This effect is mainly manifested near the triple point where the anomalous behaviour of density is observed. It had been also shown that the peculiarities of non-monotonous dependences of the specific volume on pressure and temperature within the range $1 < t < 1.2, \quad p_{tr} < p < 150\, MPa$, are in concord manifested in the behaviour of the kinematic shear viscosity.

It is very important that the shear viscosity and self-diffusion have no activation character.

It was shown that the cluster structure of water has no relation to the formation of quasi static shear viscosity and self-diffusion. The sizes of water molecules manifested in these kinetic processes are the same and this size is reduced to hard-core radius of water molecule. It is noticeably different from those values determined by the averaged interparticle potential and the equation of state.

At the same time, caloric and polarizational properties of water, first of all, the entropy, heat capacity and polarizability are determined by the thermal excitatations of clusters [45,46]. The cluster structure is also manifested in the behaviour of the structural factor [35,47].

Clusters are formed by H-bonds, therefore the distance between the nearest oxygens are more than the same distance of rotating molecules. Here it is taken into account that the tetrahedral structure of water is the most friable. Therefore the destruction of cluster structure is connected with the diminution of molecular size (see Fig.7).

## Literature

[1] D.Eisenberg and V. Kauzmann, The Structure and Properties of Water; Oxford University Press, New York, USA, 1969

[2] F. Franks, Water: A Comprehensive Treatise ; Plenum, New York, USA, 1982.

[3]. S. Magazù, G. Maisano, P. Migliardo. Hydration and transport properties of aqueous solutions of α-α-trehalose J. Chem. Phys. 109, 1170-1174 (1998); https://doi.org/10.1063/1.476662

[4]. S. Magazù, V. Villari, P. Migliardo, G. Maisano, M. T. F. Telling. Diffusive Dynamics of Water in the Presence of Homologous Disaccharides: A Comparative Study by Quasi Elastic Neutron Scattering. IV. J. Phys. Chem. B 2001, 105, 9, 1851–1855 https://doi.org/10.1021/jp002155z

[5]. Magazù, S., Migliardo, F. & Telling, M.T.F. Study of the dynamical properties of water in disaccharide solutions. *Eur Biophys J* 36, 163–171 (2007). https://doi.org/10.1007/s00249-006-0108-0

[6]. S. Magazù, F. Migliardo, M.T.F. Telling, Structural and dynamical properties of water in sugar mixtures, Food Chemistry, Volume 106, Issue 4, 2008, Pages 1460-1466, https://doi.org/10.1016/j.foodchem.2007.05.097.

[7]. S. Magazù, F. Migliardo, M. T. F. Telling. Study of the dynamical properties of water in disaccharide solutions. European biophysics journal 2007 36(2):163-171
 DOI: 10.1007/s00249-006-0108-0

[8] G.G. Malenkov. Structural and dynamical heterogeneity of stable and metastable water. Phys. A Stat. Mech. Its Appl. 314 (2002) 477-484.

[9] V.P. Voloshin, Y.I. Naberukhin. Hydrogen bond lifetime distributions in computer-simulated water. J. Struct. Chem. 50 (2009) 78-89.

[10] V.P. Voloshin, Y.I. Naberukhin, G.G. Malenkov. Percolation analysis of hydrogen bonds network in water: colouring with bond lifetime and energy. Electron. J. "Structure Dyn. Mol. Syst." 10 (2011) 12- 25, http://old.kpfu.ru/sdms/sod_10a_2011. htm.

[11] N.P. Malomuzh, V.N. Makhlaichuk, P.V. Makhlaichuk, K.N. Pankratov, Cluster structure of water in accordance with the data on dielectric permittivity and heat capacity, J. Struct. Chem. 54 (Supplement 2) 205 (2013) S24-S39, https://doi.org/10.1134/ S0022476613080039.

[12] K. Okada, M.Yao, Y. Hiejima, H.Kohno, Y. Kojihara, Dielectric relaxation of water and heavy water in the whole fluid phase J. Chem. Phys. 110 (1999) 3026 -3036, https://doi.org/10.1063/1.477897.

[13] H.R. Pruppacher. Self-diffusion coefficient of supercooled water. J. Chem. Phys. 56 (1972) 101-107, https://doi.org/10.1063/1.1676831.

[14] K. Simpson, M. Karr. Diffusion and nuclear spin relaxation in water. Phys. Rev. 111 (1958) 1201-1202, https://doi.org/10.1103/PhysRev.111.1201.

[15] L.A. Bulavin, N.P. Malomuzh, K.N. Pankratov, Character of the thermal motion of water molecules according to the data on quasielastic incoherent scattering of slow neutrons J. Struct. Chem. (Russia) 47 (2006) 54-62.

[16] L.A. Bulavin, A.I. Fisenko, N.P. Malomuz. Surprisin properties of the kinematic shear viscosity of water. Chem. Phys. Lett. 453 (2008) 183 - 187, https://doi.org/10.1016/j.cplett.2008.01.028

[17] I.V. Zhyganiuk, M.P. Malomuzh, Physical nature of hydrogen bond Ukr. J. Phys. 60 (2015) 960-974, https://doi.org/10.15407/ujpe60.09.0960.

[18] Timofeev M.V. Simulation of the interaction potential between water molecules Ukrainian Journal of Physics 61(10) (2016) 893–900, https://doi.org/10.15407/ujpe61.10.0893

[19] N.P. Malomuzh, I.V. Zhyganiuk, M.V. Timofeev, Nature of H-bonds in water vapor, J. Mol. Liq. 242 (2017) 175–180,
https://doi.org/10.1016/j.molliq.2017.06. 127.

[20] P.V. Makhlaichuk, V.N. Makhlaichuk, N.P. Malomuzh, Nature of the kinematic shear viscosity of low-molecular liquids with averaged potential of Lennard-Jones type, J. Mol. Liq. 225 (2017) 577-584, https://doi.org/10.1016/j.molliq.2016.11.101.

[21] N.P. Malomuzh, V. N. Makhlaichuk On the similarity of the self-diffusion and shear viscosity coefficients in low-molecular liquids Journal of Molecular Liquids. 295 (2019) 111729. doi:10.1016/j.molliq.2019.111729

[22] L.D. Landau, E.M. Lifshitz, Course of Theoretical Physics, Statistical Physics, vol. 5, Pergamon, Oxford, 1980 Nauka, Moscow, 1995.

[23] N.P. Malomuzh, V.P. Oleynik, Nature of the kinematic shear viscosity of water, J. Struct. Chem. (Russia) 49 (N 6) (2008) 1055–1063, http://dx.doi.org/10.1007/ s10947-008-0178-1.

[24] A.Z. Patashinskii, V.L. Pokrovskii, Fluctuation Theory of Phase Transitions, Pergamon Press, 1979.

[25] I.Z. Fisher, Statistical Theory of Liquids, University of Chicago Press, 1964.

[26] N P Malomuzh, K S Shakun Collective contributions to self-diffusion in liquids Phys. Usp. 64 (2021) 157–174 doi: 10.3367/UFNe.2020.05.038759

[27] T. V Lokotosh, N P Malomuzh, K. N. Pankratov, K. S Shakun. New Results in the Theory of Collective Self-Diffusion in Liquids. Ukrainian Journal of Physics, 60(8) (2019) 697-707. https://doi.org/10.15407/ujpe60.08.0697

[28] A. Belonoshko, T. Lukinov, J. Fu *et al.* Stabilization of body-centred cubic iron under inner-core conditions. Nature Geosci 10 (2017) 312–316. https://doi.org/10.1038/ngeo2892

[29] NIST Chemistry WebBook. SRD 69. Thermophysical Properties of Fluid Systems https://webbook.nist.gov/chemistry/fluid/

[30] V. E. Chechko, V. Ya. Gotsulskiy, N. P. Malomuzh. Surprising peculiarities of the shear viscosity for water and alcohols Journal of Molecular Liquids 318 (2020) 114096 https://doi.org/10.1016/j.molliq.2020.114096.

[31] IAPWS, Revised Release on the IAPS Formulation 1985 for the Viscosity of Ordinary Water Substance, International Association for the Properties of Water and Steam, Erlangen, Germany, 1997, 15, retrieved from http://www.iapws.org/relguide/visc.pdf.

[32] B.P. Nikolsky, Guide-Book for Chemist (in Russian).Chemistry, Moscow, 1964.

[33] J. H. Dymond , R. Malhotra . The Tait Equation: 100 Years On . International Journal of Thermophysics (1988) 9(6) 941-951. https://doi.org/10.1007/BF01133262

[34] T. Sotani , J. Arabas, H. Kubota , M. Kijima. Volumetric behaviour of water under high pressure at subzero temperature. High Temperatures - High Pressures (2000) 32 433-440.DOI:10.1068/htwu318

[35] A. K. Soper, C. J. Benmore. Quantum Differences between Heavy and Light Water. Physical Review Letters, 101 (2008) 065502.

doi: 10.1103/PhysRevLett.101.065502

[36] L. Zheng, M. Chen, Z. Sun, Hsin-Yu Ko, B. Santra, P. Dhuvad, X. Wu. Structural, electronic, and dynamical properties of liquid water by ab initio molecular dynamics based on SCAN functional within the canonical ensemble: The Journal of Chemical Physics 148 (2018) 164505 doi: 10.1063/1.5023611

[37] H. J. C. Berendsen, J.P.M. Postma, W.F. van Gunsteren et al Intermolecular Forces, edited by B. Pullman (Reidel, Dordrecht,1981), p. 331

[38] H. J. C. Berendsen, J. R. Grigera, T. P. Straatsma The missing term in effective pair potentials J. Phys. Chem. 91 (1987) 6269-6271 doi:10.1021/j100308a038

[39] Jorgensen W. L., Chandrasekhar J., Madura J. D [et al] Comparison of simple potential functions for simulating liquid water. J. Chem. Phys. 79 (1983) 926-935 https://doi.org/10.1063/1.445869

[40] J. Vrabec, J. Stoll, H. Hasse, A set of molecular models for symmetric Quadrupolar fluids, J. Phys. Chem. B 105 (48) (2001) 12126–12133, http://dx.doi.org/10.1021/ jp012542o.

[41] E. Wilhelm, R. Battino, Estimation of Lennard-Jones (6,12) pair potential parameters from gas solubility data, J. Chem. Phys. 55 (1971) 4012-4016 http://dx.doi.org/10.1063/1. 1676694.

[42] A. Beneduci, Which is the effective time scale of the fast Debye relaxation process in water? J. Mol. Liq. 138 (2007) 55-60 https://doi.org/10.1016/j.molliq.2007.07.003

[43] H. Yada, M. Nagai and K. Tanaka, The intermolecular stretching vibration mode in water isotopes investigated with broadband terahertz time-domain spectroscopy, Chem. Phys. Lett. 473 (2009) 279-283 doi:10.1016/j.cplett.2009.03.075

[44] N. P. Malomuzh, K. S. Shakun  Maxwell relaxation time for argon and water, Journal of Molecular Liquids, 293 (2019) 111413, https://doi.org/10.1016/j.molliq.2019.111413

[45] V. N. Makhlaichuk., N. P. Malomuzh.  Manifestation of cluster excitations in dielectric properties of water vapor and liquid water as well as their heat capacity. Journal of Molecular Liquids 253 (2018) 83–90 https://doi.org/10.1016/j.molliq.2018.01.018

[46] N. P. Malomuzh. Cluster structure of water and its argon-like equation of state. RENSIT, 12(1) (2020) 39-48  DOI: 10.17725/rensit.2020.12.039

[47] J. H. Eggert, G. Weck, P. Loubeyre. Structure of liquid water at high pressures and temperatures . J. Phys.: Condens. Matter 14 (2002) 11385–11394. DOI 10.1088/0953-8984/14/44/487